\documentclass{aastex6}
\usepackage{booktabs}
\usepackage{color}
\usepackage{threeparttable}

\newcommand{\niii}{N {\sc iii}] $\lambda 1750$\ }
\newcommand{\ciii}{C {\sc iii}] $\lambda 1908$\ }
\newcommand{\niv}{N {\sc iv}] $\lambda 1486$\ }
\newcommand{\siiv}{Si {\sc iv}] $\lambda 1396$\ }
\newcommand{\civ}{C {\sc iv} $\lambda 1549$\ }
\newcommand{\nv}{N {\sc v} $\lambda 1240$\ }

\newcommand{\riii}{C {\sc iii}] $\lambda 1908$/N {\sc iii}] $\lambda 1750$\ }
\newcommand{\riv}{C {\sc iv}] $\lambda 1549$/N {\sc iv}] $\lambda 1486$\ }
\newcommand{\riiib}{N {\sc iii}] $\lambda 1750$/C {\sc iii}] $\lambda 1908$\ }
\newcommand{\rivb}{N {\sc iv}] $\lambda 1486$/C {\sc iv} $\lambda 1549$\ }
\newcommand{\rheha}{He {\sc II} $\lambda 4686$/H$\alpha$\ }
\newcommand{\kms}{km~s$^{-1}$}
\newcommand{\ergs}{erg~s$^{-1}$ }

\begin{document}
\title{Carbon and Nitrogen Abundance Ratio In the Broad Line Region of Tidal Disruption Events} 
\author{Chenwei Yang\altaffilmark{1}$^,$\altaffilmark{2}}
\author{Tinggui Wang\altaffilmark{1}$^,$\altaffilmark{2}}
\author{Gary J.Ferland\altaffilmark{3}}
\author{Liming Dou\altaffilmark{4}}
\author{Hongyan Zhou\altaffilmark{5}}
\author{Ning Jiang\altaffilmark{1}$^,$\altaffilmark{2}}
\author{Zhenfeng Sheng\altaffilmark{1}$^,$\altaffilmark{2}}

%\and

%% Use the \and command so offset the last author.


\altaffiltext{1}{CAS Key Laboratory for Researches in Galaxies and Cosmology, Department of Astronomy, University of Science and Technology of China, Hefei, Anhui 230026; cwya@ustc.edu.cn}
\altaffiltext{2}{School of Astronomy and Space Sciences, University of Science and Technology of China, Hefei, Anhui 230026}
\altaffiltext{3}{Department of Physics and Astronomy, University of Kentucky, Lexington, KY 40506, USA}
\altaffiltext{4}{Center for Astrophysics, Guangzhou University, Guangzhou 510006, China; Astronomy Science and Technology Research Laboratory of Department of Education of Guangdong Province, Guangzhou 510006, China}
\altaffiltext{5}{Polar Research Institute of China, 451 Jinqiao Rd, Shanghai 200136, China}
\begin{abstract}
The rest-frame UV spectra of three recent tidal disruption events (TDEs), ASASSN-14li, PTF15af and iPTF16fnl display strong nitrogen emission lines but weak or undetectable 
carbon lines. In these three objects, the upper limits of the \riii ratio are about two orders 
of magnitude lower than those of quasars, suggesting a high abundance ratio of [N/C]. 
With detailed photoionization simulations, we demonstrate that $C^{2+}$ and $N^{2+}$ are formed in the same zone, so the C{\sc iii}]/N {\sc iii}] ratio depends 
only moderately on the physical conditions in the gas and weakly on the shape of the ionizing continuum. There are smaller than $0.5$ dex variations in the line ratio over wide ranges of gas densities and ionization parameters at a given metallicity. This allows a robust estimate of the relative abundance ratio nitrogen to carbon. We derive a relative abundance ratio of [N/C]$>1.5$ for ASASSN-14li, and even higher for PTF15af and iPTF16fnl. This suggests that the broad line region in those TDE sources is made of 
nitrogen-enhanced core material that falls back at later times. Based on stellar evolution 
models, the lower limit of the disrupted star should be larger than 0.6M$_{\sun}$. 
The chemical abundance of the line emitting gas gives a convincing evidence that the flares origin from stellar tidal disruptions.
The coincidence of the weakness of the X-ray emission with the strong broad absorption 
lines in PTF15af, iPTF16fnl and the strong X-ray emission 
without such lines in ASASSN-li14 are in analogy to quasars with and without broad absorption lines. 
\end{abstract}


\section{Introduction} \label{sec:intro}
Evidence has been mounting that in the center of most galaxies there is a 
supermassive black hole. When a star passes too close (closer than the tidal 
radius $R_{T}=r_{*}(M_{BH}/M_{*})^{1/3}$) to the supermassive black hole, it 
will be torn apart by the tidal force of the hole. Such event is called stellar 
tidal disruption event (TDE). After the disruption, about half of the stellar 
debris will be accreted onto the black hole, causing a strong flare peaking 
at soft X-ray to UV band (Rees 1988; Phinney 1989). The flare rises quickly 
on a time scale of about a month (Gezari et al. 2009, 2012; Arcavi et al. 2014), 
and falls in an approximate power-law form (Komossa\& Bade 1999; 
Gezari et al. 2015) or exponentially form (Holoien et al. 2016). Up to now, about 
60 TDE or TDE candidates were detected up to a redshift of z=0.89 (Brown et al. 2015). 
Thus, the TDE is a unique probe for the presence of supermassive black holes 
in distant quiescent galaxies, for which the radius of influence of the black 
hole is too small to be resolved by current telescopes. 
Furthermore, the energetic TDE flare will ionize the gas and heat the dust 
in the ambient environment, causing variable coronal lines (Wang et al. 2012; 
Yang et al. 2013) and transient signatures in the infrared (Jiang et al. 2016; 
Dou et al. 2016; van Velzen et al. 2016; Dou et al. 2017). Monitoring such variable features 
could constrain the gas and dust properties in the galactic nucleus. 

In many optically-bright TDEs, transient broad emission lines (BELs) were also 
detected, and these lines vanish on time scales as short as one year (e.g., Wang et al. 2011; 
Gezari et al. 2012; Brown et al. 2015). It was proposed that the lines may come from 
strong outflows launched during the early high-accretion rate stage (Strubbe \& 
Quataert 2009) or from streams of tidal debris (Bogdanovic et al. 2004; 
Guillochon et al. 2014). Although their origin has not been fully understood, 
the emission lines and their time evolution carry important information about 
the physical, dynamical and chemical properties of gas streams around the 
supermassive black hole. Such information is vital to understanding 
the formation and evolution of accretion flows and launching of outflows 
during the super-Eddington phase. 

Recently, UV spectra of three TDEs, ASASSN-14li (Cenko et al. 2016), PTF15af 
(Blagorodnova, in prep) and iPTF16fnl(Brown et al. 2017) were taken with HST. To first 
order, their UV spectra are similar to those of quasars. But a closer look reveals that 
both of them display very strong nitrogen lines while the expected carbon lines are 
relatively weak or absent. For comparison, even in one of the most nitrogen 
rich quasars Q0353-383, whose nitrogen is over abundant to oxygen by a 
factor of ~15 compared to the solar value(Baldwin et al. 2003), the \rivb and \riiib line 
ratios are smaller than those in ASASSN-14li by about one order of magnitude.
The cause of the anomalous line ratios 
is still under debate. It could be ascribed to an enhanced [N/C] abundance (Kochanek 2016a), 
collisional de-excitation of \ciii line in a dense emission 
region, or certain radiation transfer effects (Cenko et al. 2016). Enhanced narrow [N {\sc ii}] emission are reported in the ultraluminous X-ray source CXOJ033831.8-352604, which could be interpreted as origins from tidal disruption of a horizontal branch star(Irwin et al. 2010, Clausen et al. 2012).

In this work, we will use photoionization models to investigate the formation 
of the carbon and nitrogen lines, analyze the applicability of using the 
C/N line ratio as an abundance indicator, and apply the method to ASASSN-14li. 
The paper is organized as follows: in section \ref{sec:sp}, we will briefly 
summarize the observations of ASASSN-14li, PTF15af and iPTF16fnl; in section 
\ref{sec:pm} we will use the numerical photoionization code CLOUDY (Ferland et al. 
2013) to investigate the ionization structure and the formation of carbon 
and nitrogen emission lines; in section \ref{sec:result} we will inspect how the 
line ratio changes with the gas properties and give constraint of the disrupted stars; and section \ref{sec:sum} is our summary.

\section{Three TDEs with UV spectroscopy} \label{sec:sp}

Recently, three TDE, ASASSN-14li, PTF15af and iPTF16fnl, were observed during outburst with the Space Telescope Imaging Spectrograph(STIS) on HST. The detailed 
multi-band evolution and optical spectroscopic analysis of ASASSN-14li and iPTF16fnl can 
be found in Holoien et al. (2016) and Blagorodnova et al. (2017). UV spectroscopy of 
ASASSN-14li and iPTF16fnl can be found in Cenko et al. (2016) and Brown et al. (2017). In 
this section, we will briefly introduce the UV and X-ray observational properties of PTF15af 
and summarize the three objects' common features, which are relevant to our model, of the three TDEs.

\begin{figure}
\figurenum{1}
\plotone{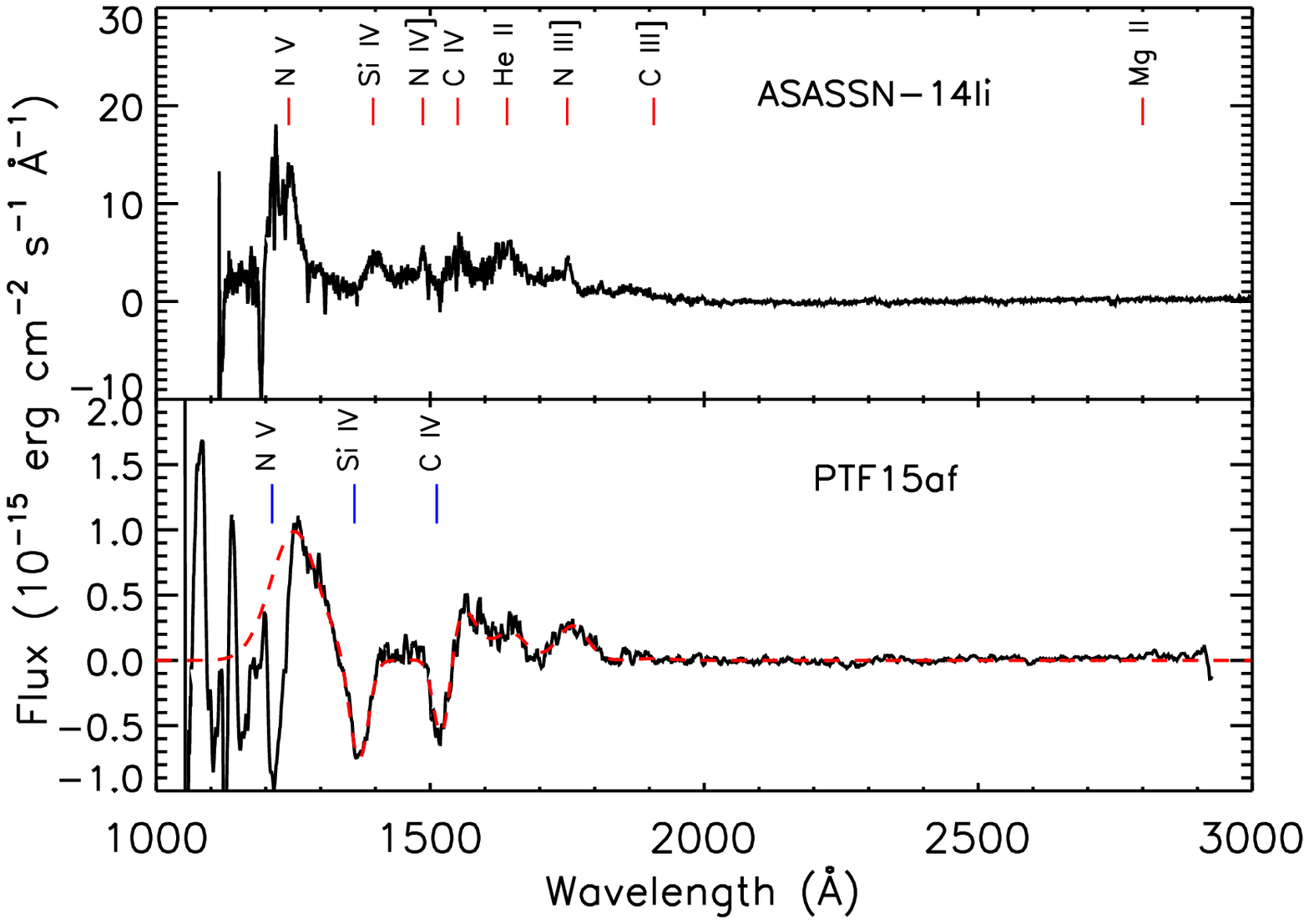}
\caption{Upper panel: HST UV line spectrum of ASASSN-14li taken $60$ days 
after its discovery. We have subtracted a blackbody of $3.5\times10^{4}K$ 
from the spectrum and mark the main broad emission lines with red lines. The \niii and \niv lines are strong while the \ciii line is absent and 
the \civ line has a strength similar to the \niv line. See Cenko et al. (2016) 
for detailed line measurements. Bottom panel: HST UV line spectrum of PTF15af after subtraction of a blackbody of $4.1\times10^{4}K$. We fit each emission and absorption line with one Gaussian model(red dash line). The \niii line is 
strong while the \ciii line is absent. Strong \nv , \siiv and \civ blue-shifted broad 
absorption lines are also detected in the spectrum and we mark them with blue vertical
 lines. The UV spectroscopic evolution of iPTF16fnl is presented in Brown et al. (2017).}\label{fig:sp}
\end{figure}

PTF15af was detected in the galaxy SDSS J084828.13+220333.4 (Blagorodnova, in prep). 
A UV spectrum was taken by HST with STIS on Mar 8, 2015 (PI:Cenko, S), 
and the emission lines look quite similar to those of ASASSN-14li (figure \ref{fig:sp}). The \nv 
and \niii emission lines are strong while \ciii is absent. Three strong broad 
absorption features around $1200\AA$, $1350\AA$~and $1500\AA$ are identified 
tentatively as blue shifted N {\sc v}, Si {\sc iv}, and C {\sc iv}, in analogy 
with broad absorption lines (BAL) quasars. The \niv emission line falls in the BAL trough of \civ, 
and the blue wing of \civ emission line is also affected by the \civ BAL, so their fluxes cannot 
be measured reliably. We fit the \niii line with a single Gaussian function after subtracting the continuum 
(see figure \ref{fig:sp}) to get the line flux. We also derive an upper 
limit for \ciii by adding a Gaussian with a width and redshift fixed at the best-fitted \niii value. The \niii emission 
line is at least one order of magnitude stronger than \ciii. Cenko et al. (2016 ) provides the line measurement of
ASASSN-14li and we fit the \ciii and \niii lines in the UV spectrum of iPTF16fnl
(t=13 days, figure 2 of Brown et al. 2017) with the same method as PTF15af.
The upper limit of the line ratios of the three objects are listed in table \ref{tab:tab1}.

We also checked PTF15af's X-ray data obtained by Swift. After PTF15af was 
detected in the UV/optical band, Swift/XRT took 14 pointed observations between 
Jan 27, 2015 and June 12, 2015 with a total exposure time of $\simeq 32.9$ ksec 
(ObsID:00033611(002-015)). We process these data following 
a standard data reduction procedure by the 'xrtpipeline' task of the software HEAsoft 
(ver.6.19) with the most updated calibration files. Unfortunately, there are too few 
photon counts to declare it be an X-ray detection, even by stacking all the 
observations together. Based on this fact, the estimated upper limit $0.3-10$ keV 
flux is $5\times10^{-15}$ erg~cm$^{-2}$~s$^{-1}$ ($90\%$, assuming a simple 
power law with a photon index~$\Gamma =2$). This value corresponds 
to an X-ray luminosity upper limit of $10^{40.9}$\ergs in the 0.3-10 keV band, making 
it a X-ray weak TDE.

All three TDEs display a strong \niii line while the \ciii line is absent. For comparison, the typical 
line ratio of \riii for quasars is $\approx40$ (Van den Berk et al. 2002) and in nitrogen-loud 
quasars the \ciii line is never absent (Jiang et al. 2008).The two TDEs with weak or no detection in the X-ray( PTF15af and iPTF16fnl) display BALs while the one with a high X-ray luminosity (ASASSN-14li) does not. We summarize the three TDEs' X-ray and BAL feature in table \ref{tab:tab1} and will discuss that in \ref{subsecion:bal}. 

\begin{table}[]
\centering
\caption{Some observation feature of the three TDEs with UV spectroscopy.}
\begin{threeparttable}
\label{tab:tab1}
\begin{tabular}{|c|c|c|c|}
\hline
object name & log(C {\sc iii}]/N {\sc iii}]) & X-ray strong & BAL\\ \hline
ASASSN-14li &$<$0.62 & yes \tnote{a} & no \\ \hline
PTF15af &$<$1.21 & no           & yes \\ \hline
iPTF16fnl &$<$0.97 & no \tnote{b} & yes \\ \hline
\end{tabular}
    \begin{tablenotes}
    	\small
        \item[a] see Holoien et al.  2016 for X-ray evolution of ASASSN-14li
        \item[b] Blagorodnova et al. 2017 find the $0.3-10keV$ luminosity of iPTF16fnl is about $2.4\times10^{39}$\ergs, four orders of magnitude lower then the peak bolometric luminosity.
    \end{tablenotes} 
\end{threeparttable}
\end{table}

\section{Photoionization Models} \label{sec:pm}

\subsection{General Considerations}
First, photoionization seems inevitable given the large luminosity in
ionizing photons produced by the TDE. Furthermore, the presence of lines from a 
wide range of ionization levels also suggests that line-emitting gas is 
photoionized with a significant column density. In order to produce strong C {\sc iv} and N {\sc iv} lines, a substantial 
fraction of C and N must be in the form of C$^{3+}$ or N$^{3+}$. These ions are 
created by electrons or photons of energies above 47.9 and 47.4 eV and 
destructed by electrons or photons of energies above 64.5 and 77.5 eV. If 
the line-emitting gas is collisionally ionized and in ionization equilibrium, 
the temperature will be between 0.9 and 1.8$\times 10^5$K according to the 
equilibrium models in the CHIANTI database (Del Zanna et al., 2015). 
At such high temperatures, only a small fraction of N is in N$^{2+}$ due to fast collisional 
ionization. So the bulk of the broad emission lines must come from photo-ionized gas. 

Photoionization models have been employed to interpret the broad line spectra 
of AGN and TDE. Hamann et al. (2002) made a detailed simulations for BLR in AGN and predicted several UV diagnostics of relative 
nitrogen abundance and overall metallicity in the gas. They adopted the AGN 
SED as the ionizing continuum and metal enrichment pattern for star-burst galaxies.
 Gaskell \& Rojas Lobos (2014) simulated photoionized gas in a TDE with CLOUDY and found that radiation transfer plays an important role in determining the observed \rheha. 
In some parameter regimes, they reproduced the observed very large 
He {\sc II}$\lambda$4686/H$\beta$ without invoking a supersolar metallicity 
of He, which was proposed originally by Peterson \& Ferland(1986). Roth et al. 
(2016) considered a model in which both continuum and emission lines 
are formed in a spherical, optically-thick envelope, whose optical depth to 
electron scattering is very large. By considering the effect of radiation 
transfer, they were able to explain the large \rheha with gas at 
the solar abundance. However, because of large optical depth, in their model 
helium is nearly fully ionized, as are metals, in order to let soft X-rays 
escape. Thus, it remains to be shown whether there are sufficient N$^{+2}$ ions in their 
model to explain the observed strong \niii while the gas is still transparent 
to soft X-rays. 

In this work we will consider models in which the lines are emitted by either 
debris streams or outflows illuminated by UV/X-ray radiation from TDE. Thus the 
emission-line region has an open geometry with a covering factor significantly 
less than unity, and lines from both sides may reach the observer. For simplicity, 
we assume a slab geometry for the emission-line cloud. We will use CLOUDY to 
simulate the emission-line spectra and compare predictions with the observed line ratios 
to constrain the relative abundance of carbon and nitrogen. 

In the condition of quasar BEL cloud, both \riii and \riv ratios are abundance indicators(Hamann et al. 2002). However, the \niv line has a much lower critical density than the \civ line, so the line ratio depends strongly on gas density. Furthermore, the \civ line may be formed in much denser gas than the \niv line. In iPTF16fnl, the \niv line appears later than the \civ line(Brown et al. 2017) and in ASASSN-14li the \civ and \niv lines show very different profile(Cenko et al. 2016). These evidences suggest that the bulk emission of the two lines may come from different regions. For these reasons, we will only not use the \riv ratio as an indicator for [C/N] to avoid 
uncertainties due to possible stratification of the emission-line region.
The \ciii and \niii lines have similar critical densities and excitation energies, and 
involve ions with similar ionization energies. In subsection \ref{subsec:struc} 
we will illustrate that the line ratio of \riii is a good indicator of [C/N].
 
\subsection{The Input SED} \label{subsec:sed}

Multi-band follow-up observations have been carried out for ASASSN-14li since its 
discovery (Holoien et al. 2016). We use the observed SED 60 days after the discovery, 
nearly simultaneous with the UV spectroscopic observation, as the first-order 
approximation to the input SED. It is composed of two black-bodies: one peaks in the 
UV with a temperature of $3.55\times 10^4$ K and a luminosity of 
$6.5\times 10^9\; L_{\sun}$; and the other peaks in the soft X-ray band with a 
temperature of $6\times 10^5$K (or 50 eV) and a luminosity of about $ 
10^{10} L_{\odot}$\ergs (Holoien et al. 2016; Miller et al. 2015). 

The ionizing continuum striking the broad line region is likely not exactly the same as 
the observed SED due to a number of reasons. First, the SED from an TDE 
may be intrinsically anisotropic, and the observed SED may include reprocessed 
emission produced by material in or outside the emission line region. Second, the observed 
SED may be affected by the extinction of the host galaxy. Finally, if the light crossing time 
of the emission line region is comparable to the evolution time scale of the flare, the lag between 
the continuum and emission line should be taken into account (Wang et al. 2012;
Saxton et al. 2016).

For these reasons, we first test whether our results are sensitive 
to the shape of the input SED. We change the relative strength of the two blackbodies 
by a factor of $4$ and find that \riiib alters by less than $10\%$ 
for the models considered in the following subsections. We also varied the 
temperature of each black body by $50\%$, and even used a standard AGN SED as 
input. We found that the resultant line ratio changed by less than $20\%$. Thus, the 
final results are insensitive to the detailed input SED. Therefore, for simplicity, we 
will only present the results for the observed SED.  

\subsection{Gas properties} \label{subsec:gas}

Next, we examine how the line ratio depends on the gas density ($n_H$), column density 
($N_H$), dimensionless ionization parameter ($U=\Phi_H/cn_H$, where 
$\Phi_H$ is the photon flux at the incident surface) and the chemical composition. We run a 
set of plane-parallel slab models covering a large parameter range: $5<\log n_H<11$~cm$^{-3}$, 
and $-1<\log U<1$. We do not consider higher densities ($n_H>10^{11}$ cm$^{-3}$) because 
\ciii and \niii lines are effectively suppressed by collisional de-excitation at those densities. 

We find that when gas density is higher than $n_H>10^{9}~cm^{-3}$, the \ciii or \niii lines 
become optically thick at high carbon/nitrogen abundances. In the TDE case, rather than static clouds,
the line emitting gas is very likely highly dynamical because the observed line profile changes with time rather 
rapidly (e.g., Holoien et al. 2016). In either the disk-wind or infalling gas-stream model,
we expect a large internal velocity dispersion within the line-emitting gas, 
so we add a turbulent velocity of 100\kms to make the relevant lines optically thin. This velocity 
is much smaller than the observed FWHMs of most of the emission lines in the UV spectra of those TDEs. 
As noted in Hamann et al. (2002), continuum pumping (resonant absorption of continuum photons followed 
by radiative decays) is insignificant for inter-combination lines with velocity dispersion smaller than 3000 \kms. 

We consider a range of gas abundances because line-emitting gas is mostly formed 
from the stellar debris, which is likely affected by stellar nuclear synthesis. 
We assume that the initial abundance of the disrupted star is similar to our Sun 
although there are suggestions that the metallicity may be higher in the centers of galactic nuclei. We 
adopt the 2010 solar composition given by Grevesse et al. (2010), $\log (C/H)=-3.57$ and $\log (N/H)=-4.17$. 

As the star evolves, the CN cycle may gradually change the C/N ratio in the hydrogen-burning core
by converting C to N while keeping the total number of C and N the same, and finally 
reach a quasi-equilibrium value of $\log (C/N)\simeq -2.2$ in about $10^8$ yr for a
one solar mass star. The equilibrium ratio increases and the time-scale shortens 
as the stellar mass increases. For stars with masses greater than 2.5 solar mass, the 
ON cycle will convert O into N, increasing further the abundance of N. 
In our work, we neglect the change of the other elements 
such as helium and oxygen which occurs over longer time scales and assume all reduced 
carbon become nitrogen ($\Delta C/\Delta N=-1$). Our grid of $\log C/N$ ranges from the solar 
value of $0.6$ to a lower value of $-2.6$ ([C/N]$=0\sim-3.2$). Note that the tidal disruption takes place so fast 
that the star may not have time to mix matter located in different radii, which fall back at different 
times, so the debris may have a wide range of C/N ratio across it.

The calculation stops when the neutral hydrogen fraction reaches $90\%$.
Because the lines used in the abundance diagnosis are formed in the hydrogen fully 
ionized zone (see below), increasing the thickness of the 
cloud further will not alter these line intensities. Also, when the strength of \niii is comparable that of \niv, as observed, the gas must be thick enough to include the N$^{2+}$ layer. 

\section{Results} \label{sec:result}
\subsection{Ionization Structure and Line Emissivities} \label{subsec:struc}
\begin{figure}
\figurenum{2}
\plotone{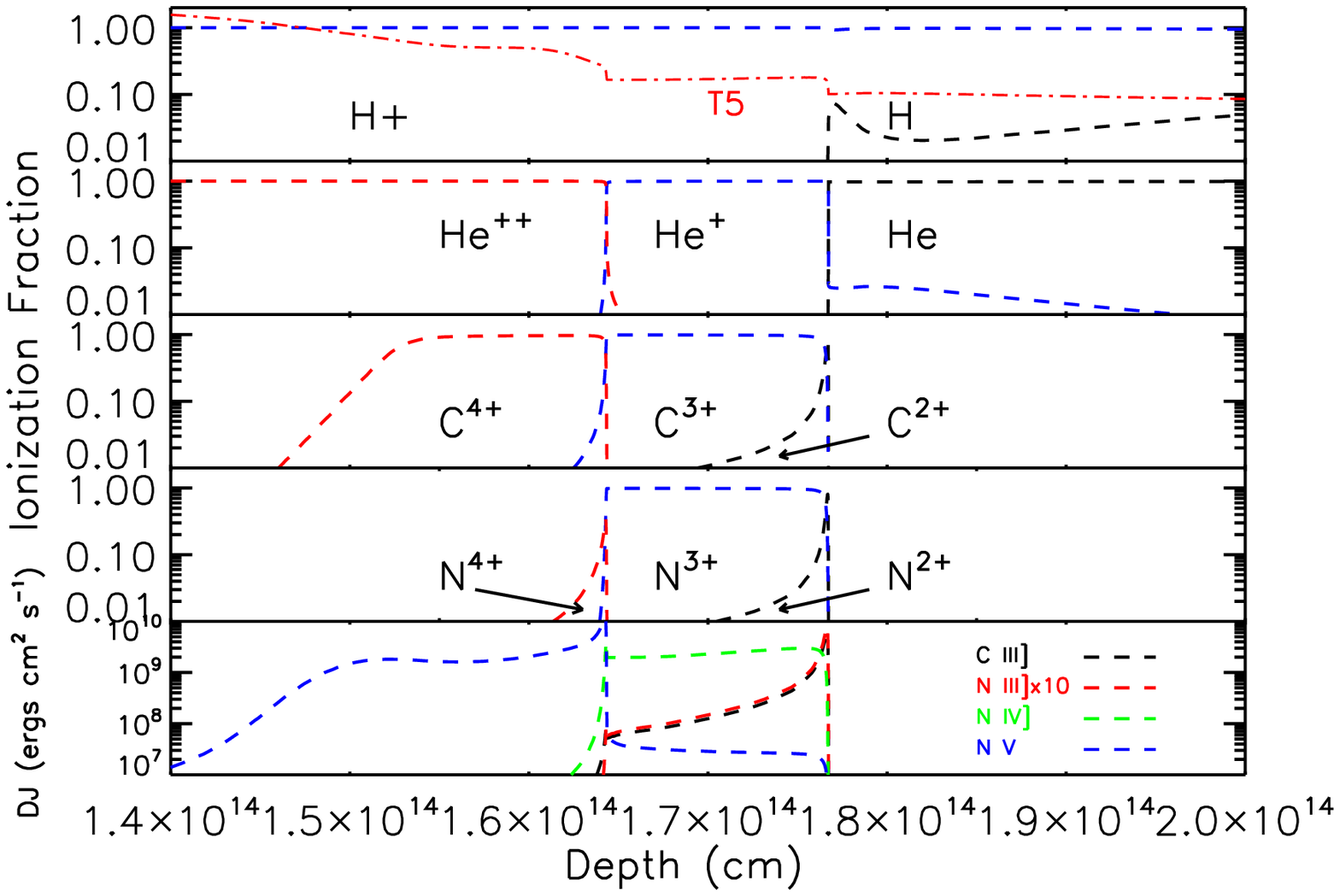}
\caption{Ionization fractions of different elements and line emissivities multiplied by spatial depth as a function of spatial depth into the cloud in one of our models ($\log U=0$, $\log n_H=10$ cm$^{-3}$, solar metallicity). From top to bottom: the ionization fraction of hydrogen, helium, carbon and nitrogen and the line emissivites of \ciii, \niv, \nv and \niii $\times10$. In the top panel we also plot the gas temperature in unit of $10^5K$. The $N^{3+} - C^{3+}$ and $N^{2+} - C^{2+}$ ions resides in co-spatial zones and the line emissivities curves of \niii line matches that of \ciii line very well. Those features are similar for other parameters. }\label{fig:str}
\end{figure}

Figure \ref{fig:str} illustrates the ionization structure of the gas in one 
of our models.
This has an ionization parameter of $\log U=0$, a gas density of $\log n_H=10$ cm$^{-3}$, and solar metallicity. In the range of ionization parameter, gas 
density and metallicity considered, the ionization structures are similar to figure 1 of Hamann et al. (2002).
The main characteristics of the ionization structure are noticed as follows:
First, ions of N and C of the same ionization state are co-spatial.
$N^{4+}$ and $C^{4+}$ reside within the $He^{++}$ zone, while $N^{3+}$, $C^{3+}$ and $N^{2+}$, $C^{2+}$ exist mainly in the $He^{+}$ 
zone. Second, the ionization fractions of $N^{3+} - C^{3+}$ and $N^{2+} - C^{2+}$ are similar at each depth, thus the absolute number densities of the pairs are proportional to 
their relative abundance. 

Next we plot the line emissivities $J$ of \nv, \niv,\niii and \ciii lines in the bottom panel of \ref{fig:str}.
Following Hamann et al. (2002), we multiply the emissivities $J$ by the spatial depth D to offset the tendency for the logarithmic scale.
For solar abundances, the \niii line is one order of magnitude weaker than the \ciii line so we multiply the \niii line emissivities by a factor of 10 to make it easier to compare.
We find that the profile of the two lines matches very well, and their ratio is proportional to the relative abundance (bottom panel of figure \ref{fig:str}).
This result is reasonable because these two lines come from ions with similar ionization structure and have similar critical density and excitation energies.
These results suggest that the pair \ciii and \niii provides a good diagnostic for the abundance ratio [C/N].
In section \ref{subsec:lr} we will check how their line ratio varies with other parameters, in addition to the [C/N]. 

\subsection{The \riii Ratio \label{subsec:lr}}
\begin{figure}
\figurenum{3}
\plotone{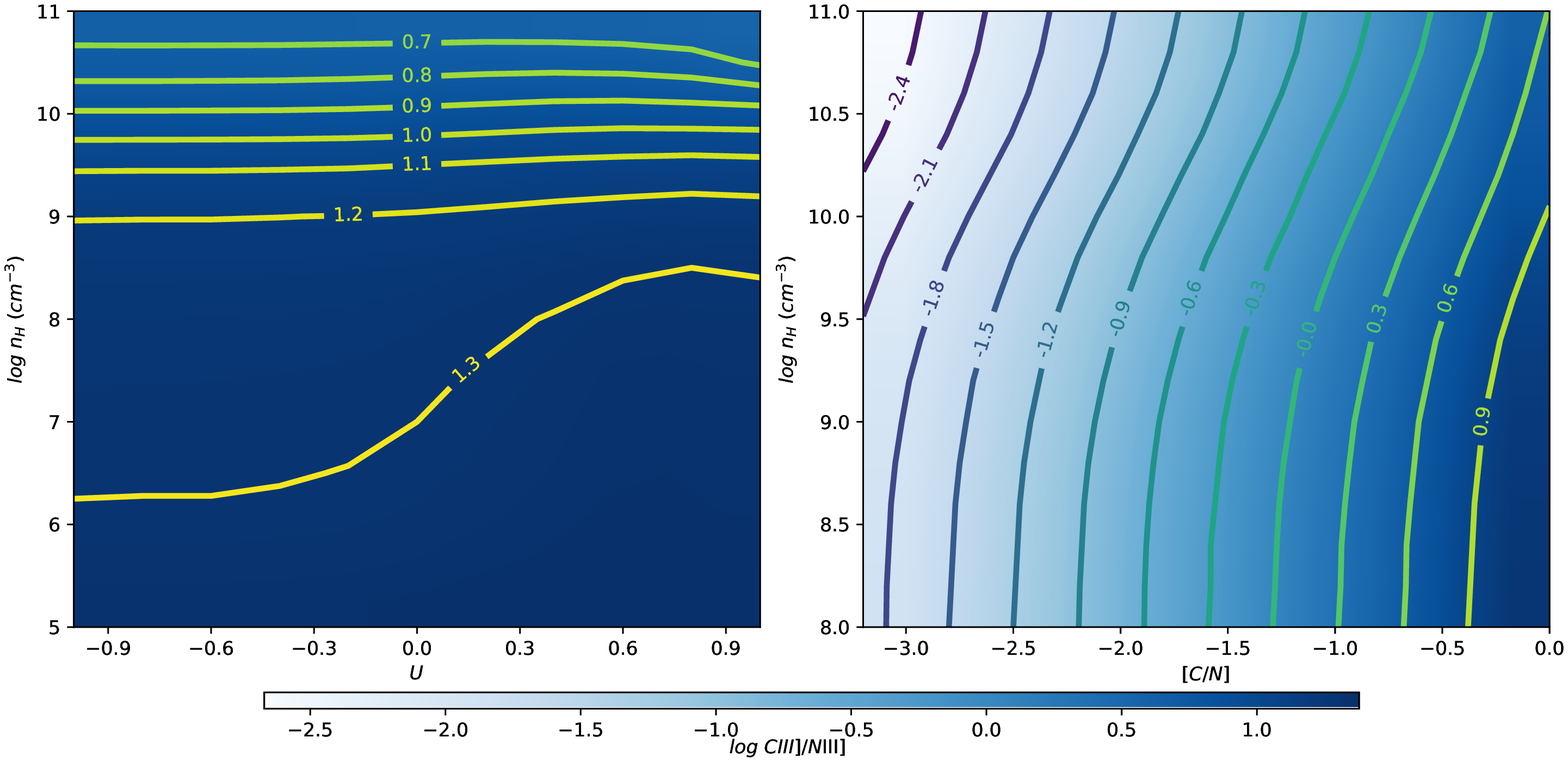}
\caption{Left panel: The logarithmic line ratio of \riii on the gas density and ionization parameter plane at solar metallicity. The line ratio has a very weak dependence on ionization parameter. When gas density increases from very low density to $\log n_H=9$, the line ratio changes about 0.1 dex.
When the gas density increases from $\log n_H=9$ cm$^{-3}$ to $\log n_H=11$ cm$^{-3}$, the line ratio varies by about 0.5 dex. 
Right panel: The logarithmic line ratio of \riii on the gas density and [C/N] plane. The \riii ratio has much stronger dependence on [C/N] than on gas density. }\label{fig:contour}
\end{figure}

In this work we consider the total emergent line fluxes produce by gas slabs. 
With the large simulations described in the last section, we first examine 
contours of the logarithmic line ratio of \riii in the ionization 
parameter vs density plane (left panel of figure \ref{fig:contour}). 
We find that at a given density and [C/N], the line ratio depends weakly 
on the ionization parameter, but vary with the gas density 
due to slightly different critical densities for these lines. When the density is 
significantly lower than the critical densities ($n_{c}\sim10^{9}$ 
cm$^{-3}$), the line ratio stays almost unchanged as the density changes. Above the 
critical densities, \riii decreases significantly as the density increases. 
When the density increases from $\log n_H=9$ cm$^{-3}$ to 
$\log n_H=11$ cm$^{-3}$, \riii varies typically by about $\sim0.6$ dex. We also check 
these results for other [C/N] values. The line ratio behaves similarly while the values at the same 
density and ionization parameter roughly scale with [C/N] value.

Since the ionization parameter has only a small effect on the line ratio, we choose $\log U=0$ as a representative value
and check how the \riii ratio varies with [C/N]. The results are plotted in the right panel of figure \ref{fig:contour}. 
We find that although the gas density affects the line ratio, the [C/N] has a much larger effect.
At a fixed gas density, when changing [C/N] from 0 to -3.2, the \riii ratio also changes by about three orders of magnitudes.

\begin{figure}
\figurenum{4}
\plotone{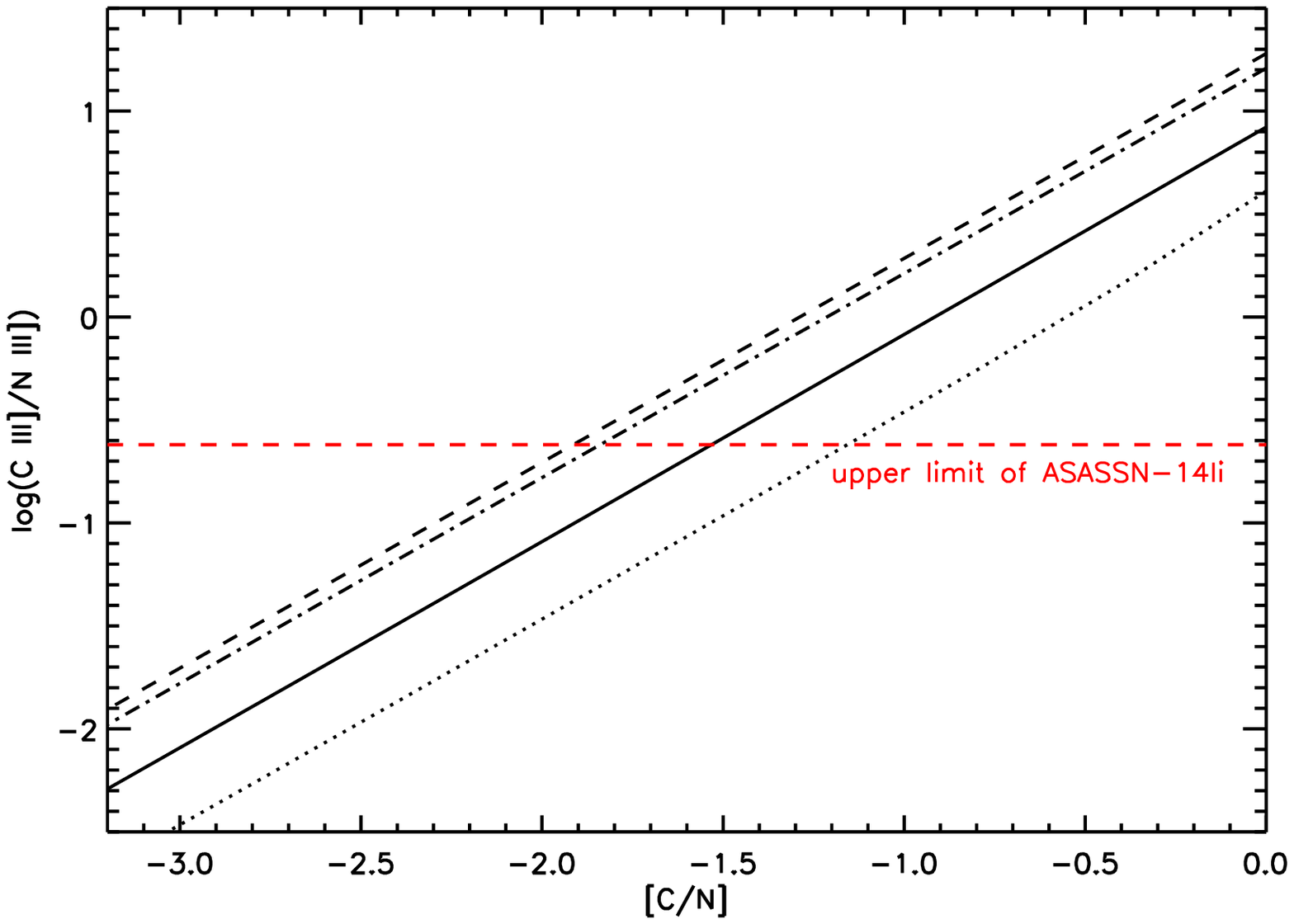}
\caption{ The predicted logarithmic line ratios of \riii versus [C/N] 
for ionization parameter $\log U=0$. The line ratio does not depend strongly on the 
ionization parameter. The dash, dash dot, solid and dotted lines represent 
$\log n_H=8,9,10$ and 11~cm$^{-3}$. The horizontal red dot line marks the observed upper limit line ratio of \riii observed in ASASSN-14li. The line ratio of \riii indicates $[C/N]<-1.5$.}\label{fig:lru0}
\end{figure}

In order to compare these predictions with observations, we plot the predicted line ratio of \riii as a function of $[C/N]$ for four densities $\log n_H=8$, 9, 10, and 11 cm$^{-3}$ 
in figure \ref{fig:lru0}. Below $\log n_H=8$, the line ratio 
does not vary much with $n_H$ while at densities higher than $10^{11}$ cm$^{-3}$ 
those inter-combination lines are effectively suppressed, leaving only very weak 
lines. Note that at $\log n_H=11$ cm$^{-3}$ and solar abundance, given the same ionizing flux $\phi_H$ and covering factor, the line intensities of \ciii and \niii are only 7\% and 11\% of those at $\log n_H=10$. 

Figure \ref{fig:lru0} shows that the observed upper-limit of \riii is too small to be 
consistent with models with solar abundance. For the density range 
considered here, the line ratio suggests an upper limit of
$[C/N] < -1.2$. As we have already discussed, it is likely that the line-emitting 
gas is distributed over a wide range in physical conditions. Because the line emissivity is 
proportional to $n^2$ until collisional excitation becomes important, the bulk of 
the lines should come from gas with a density close to the critical density ($\log n_c\sim10 $ cm$^{-3}$). 
With this assumption and using the observed line ratio from ASASSN-14li, 
we infer an $[C/N]$ of about $-1.5$, which converts to a number ratio of carbon to nitrogen of $\log(C/N)<-0.9$.
We will use that value to constrain the properties of the disrupted star in the next section.

\subsection{Constraints on the Disrupted Star, and Clues to the Nature of the Line-Emitting Gas} 
The only three TDEs with early-stage UV spectra display a high 
nitrogen to carbon line ratio, suggesting it maybe a common feature in TDEs. Based on our model, this reflects a high N/C abundance ratio in all the three TDEs. 
High N/C abundance ratios of about $\log(C/N)=-1$ have been found in several planetary nebulae (PN) (Ventura et al. 2016) and a few supernovae (SNe) (Fransson et al. 2005). If the line emitting gas origins from nitrogen enriched gas from PN or supernova remains, it's possible to produce high N/C line ratios. However, unless the PN or SNe locates quite close to the blackhole, it's difficult to explain how the nitrogen enriched gas could be accreted onto the black hole without mixing with low N/C interstellar medium which is likely massive in the galactic nucleus. So the high N/C abundance ratio suggests that the line-emitting gas is more likely composed of material from the stellar core of the disrupted star that is affected by nuclear synthesis.
As discussed in the last section, the robust upper limit of carbon to nitrogen ratio in the line emitting gas is $\log(C/N)<-0.9$, so we consider that the stellar core of the disrupted star should have N/C$\geq 10$ as an essential requirement.
In order to see what constraints can be put on the disrupted star and the nature of BLR, 
we ran a set of stellar evolution models using the code MESA (Paxton et al. 
2011). We assume that the star has a initial solar metallicity, 
$Z=0.02$. We evolve the star from the pre-main sequence, through the main 
sequence, and the model calculation stops when the hydrogen in the core 
is less than $10^{-3}$. We do not consider the sub-giant or later stages of
evolution. We use the default nuclear synthesis network for hydrogen burning 
and conventional treatment of convection. 

\begin{figure}
\figurenum{5}
\plotone{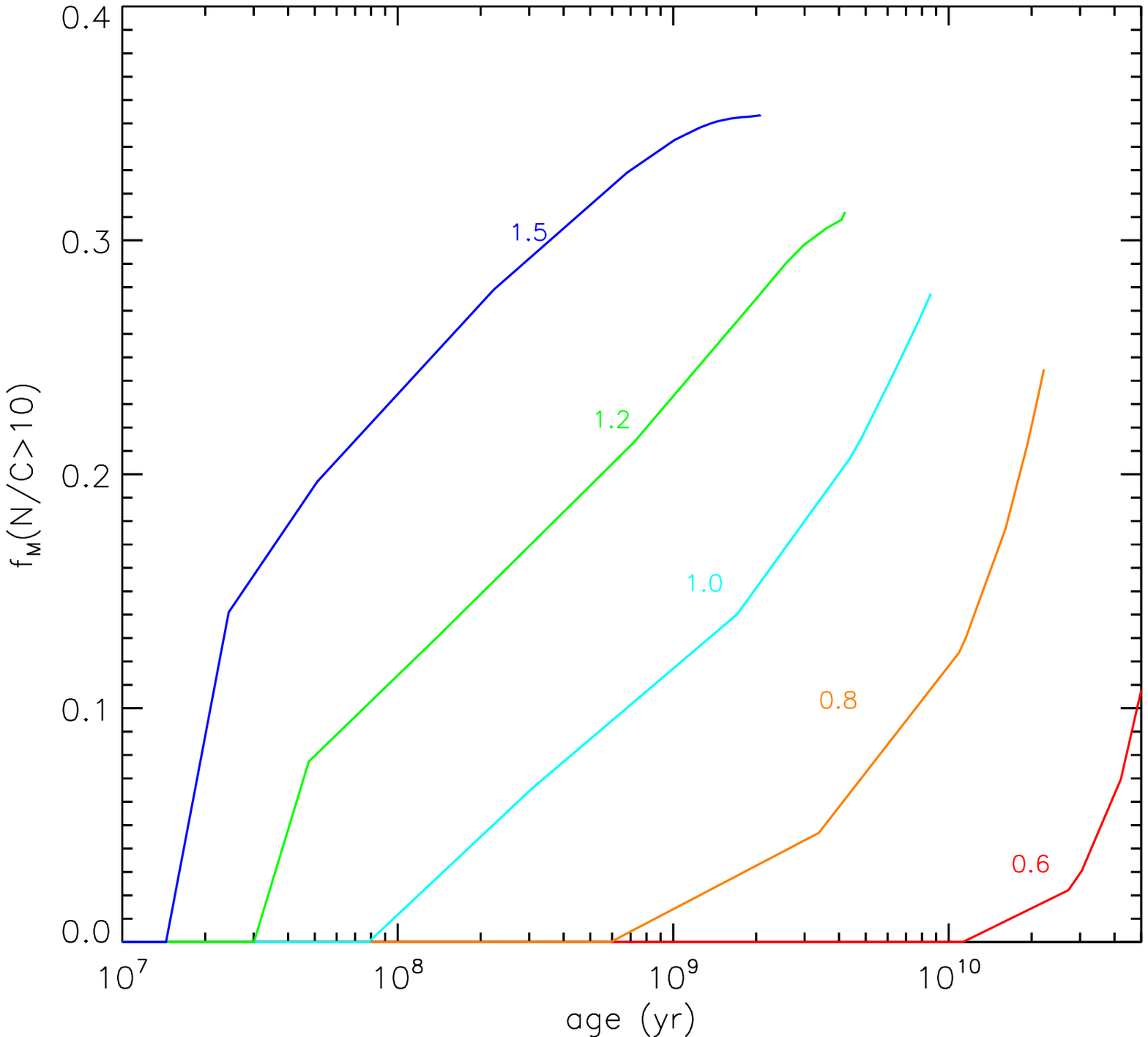}
\caption{The fraction of the central mass with N/C$\geq 10$ as a function of stellar age for stars of masses 0.6, 0.8, 1.0, 1.2 and 1.5 $M_{\sun}$. For stellar mass less than 0.6 $M_{\sun}$, the required time is longer than the age of the universe.}\label{fig:cno}
\end{figure}

Figure \ref{fig:cno} shows the fraction of the central mass with N/C$\geq 10$ as a 
function of stellar age for stars of masses 0.6, 0.8, 1.0, 1.2 and 1.5 $M_{\sun}$. 
Up to 35\% mass of the stellar core can have such a high C/N ratio during the main 
sequence for those stars, and the fraction decreases with stellar mass and the stellar 
age. The host galaxies of the three TDEs can be classified as E+A galaxies by their strong H$\delta$ absorption lines
(French et al. 2016; Blagorodnova et al. 2017; Brown et al. 2017) and there is evidence that TDEs are preferentially found in E+A galaxies (Arcavi et al. 2014; French et al. 2016).
If the disrupted stars were born during the last burst in an E+A host, their ages are about $10^8$ yr. According to figure \ref{fig:cno}, only stars with a mass of 1 solar or above possesses a core with N/C$\geq 10$. At stellar masses less than 0.6 $M_{\sun}$, 
it requires a time longer than the age of the universe for the star to evolve to the stage 
where their core reaches a sufficiently high N/C ratio. Thus, we can safely say that the 
disrupted star must be larger than 0.6 $M_{\sun}$. In contrast, Kochanek (2016b) 
suggested that most disrupted stars have a mass around 0.3 $M_{\sun}$. The three 
objects show very low \riii, requiring a considerably larger stellar mass. 
To reconcile their results with these observations, it may be that initial mass function (IMF) of stars around a galactic nucleus does 
not extend to very low stellar mass. Since we have only three cases, it may also be a coincidence. But future UV spectroscopic observations can certainly 
provide important constraints on this questions. 

When a star is disrupted, about half of the debris is bound to 
the black hole while the other half remains unbound. The debris from the 
stellar core will have an energy close to zero (e.g., Lodato \& Rossi 2011), 
that separates the bound and the unbound debris. 
Such debris eventually falls back into the black hole and forms the 
slowest unbound stream. For some reason, this core material makes a major contribution 
to the broad emission lines, thus it must subtend a substantial solid 
angle relative to the continuum source. Unfortunately, theoretical work suggests 
that the orbits of debris align for a non-rotating blackhole, thus, the covering factor of bound or 
unbounded debris will be small (Kochanek 1994). However, massive outflows 
are expected to launch during the early phase super-Eddington accretion 
(e.g., Jiang et al 2015). We conjecture that the infalling debris 
 may be interacting with these outflows at late times, spreading over a wide 
solid angle. The gas is then illuminated by the TDE flare and emits the broad 
emission lines. 

\subsection{Analogy to BAL Quasars}\label{subsecion:bal}

It has been known that some UV and optically discovered TDEs are X-ray weak although all 
early discoveries of TDEs were made in the X-ray band (e.g. Grupe et al. 1995; Bade et al. 1996). With an increasing number of TDE events,
there seems to be two distinct populations with regard to their X-ray properties: X-ray strong and X-ray weak.
The continuum and emission line properties for the two populations looks remarkably similar, suggesting the central engine is the same.

PTF15af and iPTF16fnl are X-ray weak and their UV spectra show BALs, while 
ASASSN-14li is X-ray bright and its UV spectrum does not show BALs. We note that Mg{\sc II} 
and Fe{\sc II} BALs were reported very recently in the TDE candidate PS16dtm with a NLS1-like 
spectrum, and it is also X-ray weak (Blanchard et al. 2017). In the literature, we found that another TDE PS1-11af also displayed Mg{\sc II} BAL (Chornock et al. 2014). Unfortunately, there is no X-ray follow-up observation. Given the very small sample of TDE, it is possible that X-ray weakness and BAL are only a coincidence. However, this relation is remarkable similar to that observed in quasars: BAL quasars are X-ray weak while non-BAL quasars are mostly X-ray strong. The analogy suggests that it may be real. Future spectroscopic follow-up in UV and X-ray are essential to confirm or reject this. 

In quasars, it is generally accepted that BAL outflows cover only a fraction of sky. When the line of 
sight passes through outflows, we detect a BAL QSO, otherwise we see a non-BAL QSO. The covering 
factor of outflows increases with accretion rate and maybe also with the black hole mass(Ganguly et al. 2007). One argument for this is the remarkable similarity in the properties of emission lines and UV to optical continuum for BAL and non-BAL QSOs. As we mentioned, we do not see any distinction between the emission line and UV/optical continuum slope between TDE with or without BALs, indicating that a similar scenario may work for TDE as well. In particular, soft X-rays are required to create $N^{+4}$, so the strong NV emission line in those objects indicates strong soft X-ray emission off our line of sight. 

Physically, the connection between BALs and X-ray weakness has been studied for many years 
in quasars, but still no consensus has been reached. It is commonly assumed that strong soft X-ray emission from quasars is filtered by
a highly-ionized thick shielding gas to keep the gas at proper ionization level to ensure effective radiative acceleration (Murray et al. 1995). 
There is evidence for both X-ray absorption and intrinsically weak X-ray emission (Fan et al. 2009; Luo et al. 2014). 
It is also controversial whether X-ray shielding is necessary for gas acceleration (e.g. Hamann et al. 2013). 
It is also proposed that the outflow itself may be highly ionized and massive, so it can produce soft X-ray absorptions (Wang et al. 2000). 
This may be particularly true for TDE, whose X-ray spectra are generally very soft, thus a rather moderately thick gas will produce strong absorption. Future UV 
spectroscopic monitoring of such objects will help to understand the physical process of formation and evolution of outflows by study the variation of the absorption lines.

\section{Summary} \label{sec:sum}
In this work we use photoionization models to calculate the \riii line ratios in the framework of a TDE. 
We find that reasonable ranges of the ionization parameter and SED shape have only weak effects on the line ratio. The gas density can affect line ratios by $\sim0.5$dex. The very small \riii line ratio can be naturally explained by the small relative abundance ratio of carbon to nitrogen. In our model calculations, the line ratio of \riii is a good indicator of the relative abundance of carbon to nitrogen. Using the upper limit of the line ratio observed in ASASSN-14li, we estimate the upper limit of carbon to nitrogen to be $[C/N]<\sim-1.5$ or $\log(C/N)<-0.9$. So the broad emission lines may come from gas which originated in a stellar core in which the CNO cycle has converted most carbon to nitrogen and helium fusion has not begun to increase the carbon abundance. Based on stellar evolution models, we estimate the mass of the disrupted star to be at least 0.6M$_{\sun}$. The abnormal abundance provides ever strong evidence for TDEs. We also note a potential connection between the appearance of BALs and X-ray weakness in TDEs.

\acknowledgments
We acknowledge the financial support by the Fundamental Research Funds for the Central Universities, the Strategic Priority Research Program “The Emergence of Cosmological Structures” of the Chinese Academy of Sciences (XDB09000000), NSFC (NSFC- 11233002, NSFC-11421303, U1431229) and National Basic Research Program of China (grant No. 2015CB857005).
GJF acknlowledges support from NSF (1412155), NASA (ATP13-0153), and STScI (HST-AR- 13245, GO-12560, HST-GO-12309, GO-13310.002-A, HST-AR-13914, HST-AR-14286.001 and HST-AR-14556).

\end{document}